# Ultrastable low-noise current amplifier

Dietmar Drung[1], Christian Krause[1], Ulrich Becker[2], Hansjörg Scherer[2], and Franz Josef Ahlers[2]

[1]Physikalisch-Technische Bundesanstalt (PTB), Abbestraße 2-12, 10587 Berlin, Germany

[2]Physikalisch-Technische Bundesanstalt (PTB), Bundesallee 100, 38116 Braunschweig, Germany

**Abstract**

An ultrastable low-noise current amplifier (ULCA) is presented. The ULCA is a non-cryogenic instrument based on specially designed operational amplifiers and resistor networks. It involves two stages, the first providing a 1000-fold current gain and the second performing a current-to-voltage conversion via an internal 1 MΩ reference resistor or, optionally, an external standard resistor. The ULCA's transfer coefficient is extremely stable versus time, temperature and current amplitude within the full dynamic range of ±5 nA. A low noise level of 2.4 fA/√Hz helps to keep averaging times low at small input currents. A cryogenic current comparator is used to calibrate both input current gain and output transresistance, providing traceability to the quantum Hall effect. Typically, within one day after calibration, the uncertainty contribution from short-term fluctuations of the transresistance is below one part in $10^7$. The long-term stability is expected to be better than one part in $10^5$ over a year. A high-precision variant is available that shows a substantially improved stability at the expense of a higher noise level of 7.5 fA/√Hz. The ULCA also allows the traceable generation of small electrical currents or the calibration of high-ohmic resistors.

**1.      Introduction**

The use of single-electron transport (SET) devices is a unique and elegant approach for the realization of the future quantum-based SI unit ampere [1, 2]. Recent advances in this field enable the generation of direct currents with uncertainties of one part per million (ppm) or better [3, 4]. However, the current generated by a SET device is typically limited to about $10^9$ electrons per second (160 pA). The traceable precision measurement of sub-nA direct currents with room temperature equipment is currently limited to uncertainties of about 10 ppm [5]; an exceptionally low value of 1 ppm was demonstrated with a very sophisticated setup [3]. Up to now, lower uncertainties of the order of 0.1 ppm at 100 pA are believed to be possible only with a cryogenic current comparator (CCC) [6]. CCCs with over 10000 turns were developed for the amplification of currents from SET devices (for example [7, 8]).

Basically, a CCC compares two currents with extremely high accuracy [6]. To suppress offset and drift effects, the currents are periodically reversed (or switched on and off if reversal is not possible). The repetition frequency $f_R$ of the single measurement (two current reversals) is typically below 0.1 Hz. The CCC involves a superconducting quantum interference device (SQUID) serving as a null detector. The SQUID has a strongly nonlinear, periodic voltage-versus-flux ($V$-$\Phi$) characteristic, with the period being the flux quantum $\Phi_0$. In a typical resistance comparison, quantum Hall resistance (QHR) versus 100 Ω standard resistor, the two currents through the CCC windings produce opposite fluxes of about $10^4$ $\Phi_0$ in the SQUID [9]. At this high flux level, an uncertainty of the result of around one part in $10^9$ is routinely achieved, corresponding to an equivalent uncertainty of the zero-flux level in the order of $10^{-5}$ $\Phi_0$.

Applying a CCC to the measurement of a 100 pA direct current lowers the flux linkage drastically; now, the equivalent uncertainty of the zero-flux level should be as low as a few $10^{-8}$ $\Phi_0$ in order to achieve 0.1 ppm overall uncertainty, even when using a CCC with ten thousands of turns. Due to the nonlinear $V$-$\Phi$ characteristic, dc flux shifts can occur by mixing down of rf interference, e.g., from the equipment used to drive the SET current source under test. In this case, systematic errors may occur which cannot be evaluated in practice due to the long averaging times required to achieve a sufficiently low random uncertainty. As an example, a CCC-based current amplifier is considered that involves a SQUID with a sinusoidal $V$-$\Phi$ characteristic and a working point at the region with maximum slope $\partial V/\partial \Phi$ just 0.01 $\Phi_0$ away from the inflection point. It is further assumed that a sinusoidal rf flux of 0.01 $\Phi_0$ amplitude is applied to the SQUID and that the rf amplitude differs by only 1% at both polarities of the CCC's input current. In this example, a systematic zero-flux error of $2\times10^{-7}$ $\Phi_0$ will occur [10] that is about an order of magnitude higher than the acceptable level discussed above.

Recently, the systematic uncertainty due to down-mixing of high-frequency interference was exemplarily investigated with a 14-bit CCC in a ratio-error test configuration [8]. The battery chargers of the current sources in the resistance bridge, operating at a base frequency of about 180 kHz, generate noticeable rf interference to the system and, thus, could be used deliberately as rf sources. By turning the chargers on and off, the rf level coupled to the SQUID was changed, yielding a reproducible dc shift in the SQUID output and an apparent ratio error. To demonstrate that the observed effects were indeed caused by down-mixing, the working point of the SQUID was varied in a certain range. With the chargers turned off, the systematic zero-flux error was typically $10^{-7}$ $\Phi_0$. However, when the chargers were turned on, the error increased considerably (up to $7\times10^{-7}$ $\Phi_0$) and a clear dependence on the SQUID working point was observed. Furthermore, the error was not stable in time (on a scale of weeks) and changed after refilling the liquid helium container of the CCC dipstick.

To avoid these potential systematic errors in CCC-based current amplifiers, we have developed an ultrastable low-noise current amplifier (ULCA) as an alternative [11, 12]. The ULCA is a room-temperature instrument with op amps and resistor networks. Its transfer coefficient is extremely stable and it is well suited for the measurement of sub-nA currents. The ULCA is traceably calibrated with a CCC at high currents of about ±10 nA where nonlinear effects in the SQUID are sufficiently low.

**2.    ULCA design and application**

The ULCA is a very versatile instrument. This chapter gives a detailed description of the fundamentals required to understand its operation and performance. The basic configurations for the different applications are presented as well as special setups for traceable calibration and self-test. The ULCA's experimental performance is demonstrated in chapter 3 by a selection of representative experiments.

*2.1.    Basic concept*

The ULCA concept is schematically illustrated in figure 1. The ULCA consists of two stages, the first providing a 1000-fold amplification of the current $I_{IN}$ from a device under test (DUT) and the second performing a current-to-voltage conversion. The operational amplifiers OA1 and OA2 in figure 1 are sophisticated circuits comprising several monolithic op amps. This results in a low input noise and a very high overall open-loop gain of well above $10^9$. The ±44 V output stage of the first amplifier OA1 is realized by using discrete transistors; the required ±47 V supply is generated from the ±5.5 V main supply voltage via a homemade, low-power charge pump. To minimize power line interference, the ±5.5 V supply is energized from a single 12 V lead battery housed in a remote battery box. Each ULCA channel involves two batteries, one of which is charged while the other is used for powering the amplifier. This guarantees uninterruptible battery operation and allows continuous measurements over arbitrarily long time intervals.

The total supply current drawn from the battery is as low as 11 mA. This results in low power consumption and keeps the internal temperature close to ambient temperature. The ULCA is available in single-channel or dual-channel variants. Both channels in a dual-channel ULCA are completely independent from each other, including separate batteries and temperature sensors. Besides minimizing power on the ULCA printed-circuit board (PCB), special care was taken to keep the power as constant as possible, i.e., independent of the output signal, to avoid nonlinearities from thermal interaction between output and input stages of the amplifiers OA1 and OA2. This ensures that the high overall gains calculated from the individual gains of the amplifier stages are indeed achieved in practice. Consequently, the transfer coefficient of the ULCA should be constant over the full dynamic range with negligible influence from the op amps. Anomalies near zero input like those observed in [13] should not be intrinsic to the ULCA.

In normal operation, the ULCA effectively acts as a current-to-voltage converter. Its overall transresistance is $A_{TR} = G_I R_{IV}$ where $G_I = 1000$ is the current gain of the input stage and $R_{IV}$ is the current-to-voltage coefficient of the output stage. The latter is practically equal to the feedback resistance of the output stage thanks to the high open-loop gain of OA2. In voltage output mode (output = VOUT), the internal 1 MΩ resistor is used for feedback, yielding $A_{TR} = 1$ GΩ. In current output mode (output = IOUT), an external standard resistor $R_{ext}$ is applied. With the actual amplifier design $R_{ext}$ can be chosen between zero (for example a CCC winding during calibration) and 100 MΩ. This way, the transresistance can be increased up to 100 GΩ, or the performance can be improved if the quality of the external standard resistor exceeds that of the internal 1 MΩ metal-foil resistor.

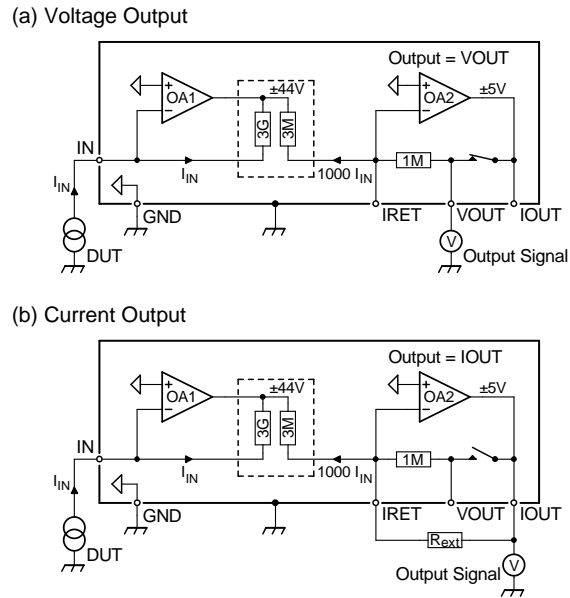

**Figure 1.** Basic schematics of the ULCA. (*a*) Voltage output, (*b*) current output. The input current $I_{IN}$ is amplified 1000 times by operational amplifier OA1 and a $3\,G\Omega/3\,M\Omega$ resistor network. The output stage OA2 converts the amplified current into a voltage via the internal $1\,M\Omega$ reference resistor (output = VOUT) or an external standard resistor $R_{ext}$ (output = IOUT), respectively. The output voltage range is ±44 V for OA1 and ±5 V for OA2. The internal reference potential (open triangles) is connected to the ULCA's metal housing via a short on the GND connector.

At low values of $R_{ext}$ the cable resistance between the ULCA and the external standard resistor has to be considered. A four-terminal measurement may be required that can be realized by connecting the current-carrying wires of the standard resistor to OA2 (IOUT and IRET) and the voltage-sensing wires to the voltmeter. However, in this configuration the voltmeter is connected in parallel to $R_{ext}$ and its input resistance changes the overall transresistance accordingly. This can be practically avoided if the low-potential side of the voltmeter is connected to ground as shown in figure 1(*b*), and another voltmeter is used to measure the small voltage difference between the low-potential side of the standard resistor and ground. It can be shown that the input resistance of the voltmeter at the output of OA2 has no effect due to the circuit design, and that the input resistance of the voltmeter at the standard resistor's low-potential side is strongly suppressed because of the low voltage across the voltmeter input. Furthermore, the requirements on accuracy (and cost) of this voltmeter are considerably relaxed compared to the instrument at the amplifier output that directly affects the measurement uncertainty.

A key component in the ULCA concept is the $3\,G\Omega/3\,M\Omega$ resistor network at the output of OA1. In the following, it will be simply referred to as "3 G$\Omega$ network" because the input resistor networks of all ULCAs implemented so far provide a 1000-fold current gain. For highest precision, the network is realized with NiCr thin-film resistor technology; thick-film resistors have inadequate stability and exhibit a too high voltage dependence of resistance. In the original approach [11], a large number of integrated $10\,M\Omega/10\,k\Omega$ matched resistor pairs in SO-8 package were considered (SO series from Powertron / Vishay Precision Group). The individual resistor pairs are fabricated on a common substrate and, consequently, are highly matched and thermally well coupled. In spite of a careful chip layout, tiny systematic differences remain between the high-ohmic and low-ohmic sides. This results in a non-vanishing, slightly positive temperature coefficient of $G_I$ (about +1 ppm/K for the latest generation) that can be compensated by adding a series resistance with appropriate temperature dependence to the low-ohmic side (e.g., a number of PT100 temperature sensors). However, it remains an open question whether the systematic differences also affect the long-term stability of $G_I$.

To minimize systematic differences between both sides of the network, an alternative approach was developed. The network is now implemented by about 3000 identical 2 MΩ thin-film chip resistors (0805 size = 2 mm × 1.25 mm). All resistors of the network are taken from the same fabrication lot to maximize matching. The base element consists of 31 resistors in series at the high-ohmic side and 32 resistors in parallel at the low-ohmic side. A total of 48 base elements are connected in series in a meander-like arrangement [11], yielding total resistances of 3 GΩ and 3 MΩ for the two sides of the network, respectively. To increase the resistance ratio from 31×32 = 992 to 1000, a total of 24 resistor pairs (each connected in parallel) are added on the high-ohmic side. Two adjacent base elements share one resistor pair in order to obtain a regular distribution. A high flexibility is attained because the resistor grade, value or manufacturer can easily be substituted as long as the case size is not changed. For example, the 3 GΩ networks involve 2 MΩ resistors with ±0.1% tolerance and ±25 ppm/K temperature coefficient. Lowering the resistance to 200 kΩ and keeping the 0805 size, ±0.01% tolerance and ±10 ppm/K temperature coefficient are available. The resulting 300 MΩ networks have substantially improved stability at the expense of an increased noise level (see chapter 3).

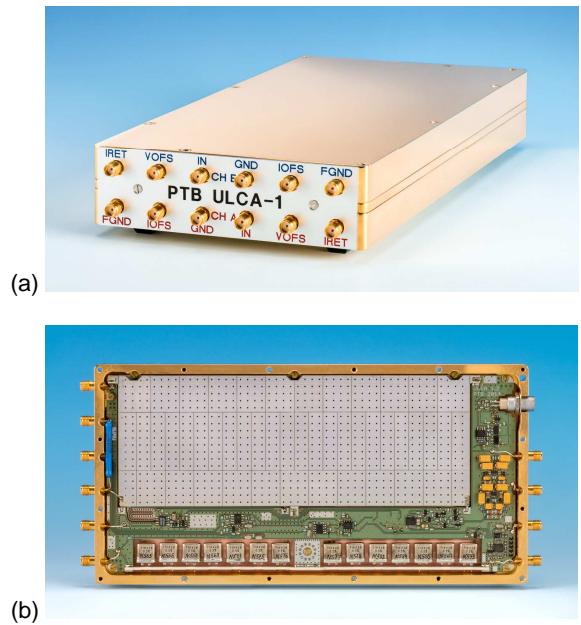

(a)

(b)

**Figure 2.** Photographs of a ULCA-1 prototype. (*a*) Complete two-channel unit with a total size of 23 cm × 12 cm × 4 cm (without connectors) and a mass of about 3½ kg. The housing is milled from massive copper bars to maximize thermal stability. All connections to inputs and outputs of the instrument are made with SMA connectors. (*b*) Single ULCA channel without cover plate. The resistor network for the 1000-fold current amplification is located under the silvery guard board. The metal-foil resistor array for the current-to-voltage conversion is visible at the bottom corner of the box.

Figure 2 shows photographs of a two-channel ULCA-1 prototype (cf. section 2.6). For maximum thermal stability, the housing is milled from massive Cu. The 3 GΩ network is covered by top and bottom guard plates (see section 3.2). At the bottom corner of the copper housing in figure 1(*b*), the internal 1 MΩ standard resistor is visible. It consists of 14 hermetically-sealed bulk metal-foil resistors (selected set of VHP101 from Vishay Precision Group). A modular PCB construction was chosen, i.e., both the 3 GΩ network and the 1 MΩ standard resistor can easily be replaced and tested separately before assembling.

*2.2.    Current generation with ULCA*

The ULCA was originally intended only for the *measurement* of sub-nA currents. Recently it was upgraded with an additional option for the *generation* of small electric currents. Figure 3 depicts both options, i.e., current measurement (input = AMP) and current generation (input = SRC). For simplicity, the voltage output mode is shown only; current output may alternatively be used according to figure 1(*b*). Compared to figure 1, two connectors were added (TEST and FGND) and a switch was inserted between the output of OA1 and the high-potential side of the 3 GΩ network. For current measurement, this switch connects the output of OA1 with the network, i.e., nothing is changed compared to figure 1. For current output, however, the output of OA1 is connected to external ground (the ULCA's copper housing) and an appropriate voltage from a function generator is applied to the 3 GΩ network. This voltage causes an output current $I_{OUT}$ flowing through the high-ohmic side of the 3 GΩ network and via the IN connector into the DUT, for example a picoammeter under calibration. The 1000-fold output current is passed through the 1 MΩ feedback resistance of OA2 and a proportional output voltage is metered with a traceably calibrated voltmeter.

Ideally, the DUT acts as a short for the current source. In practice, however, a small burden voltage appears across the DUT due to its finite input resistance. The effect of this burden voltage on the measurement result is eliminated in figure 3(*b*) by OA1 providing a virtual ground at the IN connector; the potential at the IN connector is thus always equal to the internal reference potential (open triangles). Therefore, the current flowing into the output stage is exactly proportional to the current through the DUT, with a gain equal to the inverse resistance ratio of the 3 GΩ network. As the low-potential side of the voltmeter is connected to the internal reference potential (and not to the housing), the overall transresistance in SRC mode is identical to that obtained in AMP mode, $A_{TR}$ = 1 GΩ. Note that the amplitude of the output current depends slightly on the burden voltage. This is, however, uncritical because the voltmeter always displays the correct current value equal to amplitude multiplied by $A_{TR}$ = 1 GΩ. The basic idea is that the current can be generated with modest precision if it is monitored at the highest possible accuracy.

A picoammeter can be calibrated with the inherent ULCA accuracy by comparing the currents displayed by the DUT (picoammeter) and the voltmeter. If the DUT and the voltmeter are read out simultaneously, even dynamic effects can be efficiently suppressed, e.g., settling effects after current reversal or low-frequency amplitude fluctuations of the function generator. This relaxes the demands on the function generator considerably. High precision and traceable calibration of the voltmeter are essential, but the function generator does not need calibration. The SRC mode is not only useful for the calibration of picoammeters, but has also practical relevance for consistency tests of the ULCA function: In a two-channel ULCA, one channel may be set into SRC mode to generate a test current for the other channel set into AMP mode. Interchanging the channels then allows mutual consistency tests. Furthermore, a ULCA with a more stable 300 MΩ network may be used to calibrate a less accurate 3 GΩ unit.

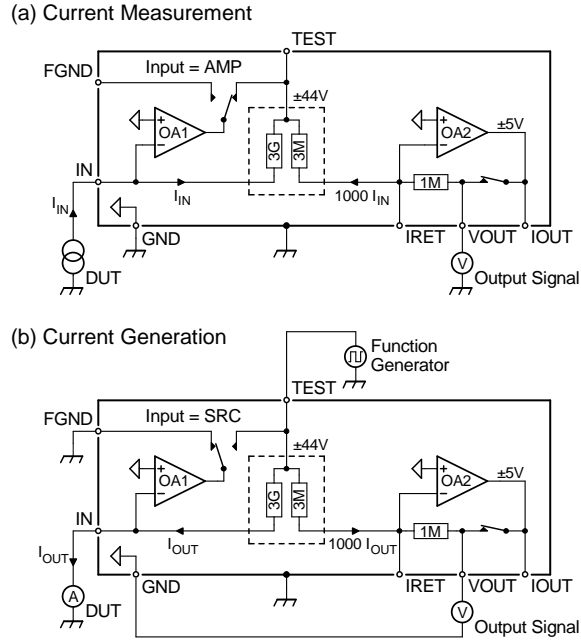

**Figure 3.** Basic operation modes of the ULCA with voltage output. (*a*) Amplifier mode for current measurement, (*b*) source mode for current generation. In source mode (input = SRC) the IN connector is used for the output current $I_{OUT}$. Amplifier OA1 forces the voltage at IN to internal ground potential. This ensures that the burden voltage at the DUT does not affect the displayed current measured with a voltmeter between the VOUT and GND connectors.

*2.3.    ULCA self-test configuration*

In the SRC mode described above, the accuracy of the voltmeter reading directly enters the overall measurement uncertainty. For the investigation of the input current gain $G_I$, the most crucial component in the ULCA concept, the demands on the voltmeter performance can be substantially reduced by using a special self-test configuration depicted in figure 4. One ULCA is set into SRC mode to generate the input current for the other ULCA (cable between IN connectors). In addition, the 1000-fold amplified input current is passed from the current-generating device into the current-measuring unit (cable between IRET connectors). Ideally, the currents cancel each other at the output stage resulting in zero output. In practice, a small imbalance occurs and the voltmeter displays a weak error signal from which the difference in the relative current gain deviations of the two resistor networks can be determined. The input current noise is increased because Nyquist noise of both resistor networks contribute.

The described self-test is intended for *in situ* characterization of resistor networks at highest precision but with modest test equipment. The output amplifier OA2 of the current-generating ULCA has to be disconnected from the low-ohmic side of the resistor network, which can be done electronically by setting the output selector into OFF position. The self-test option is a very helpful tool for long-term analysis of the input current gain (section 3.7). It can also be used to investigate the linearity of the resistor networks at low input currents. For this application, it should be possible to pass unequal currents through the two networks because otherwise systematic nonlinear effects might cancel each other at the output.

Therefore, a 10 GΩ thick-film resistor was added at the input of OA1, and an extra 9 MΩ resistor was connected between the VOUT and IOUT connectors when the output stage is disconnected (output = OFF in figure 4). Together with the internal 1 MΩ metal-foil resistor, a total resistance of 10 MΩ is present between IOUT and IRET. Applying a voltage to the connectors IOFS and IOUT of the current-generating ULCA in figure 4 generates the required offset currents $I_0$ and $1000\,I_0$, respectively.

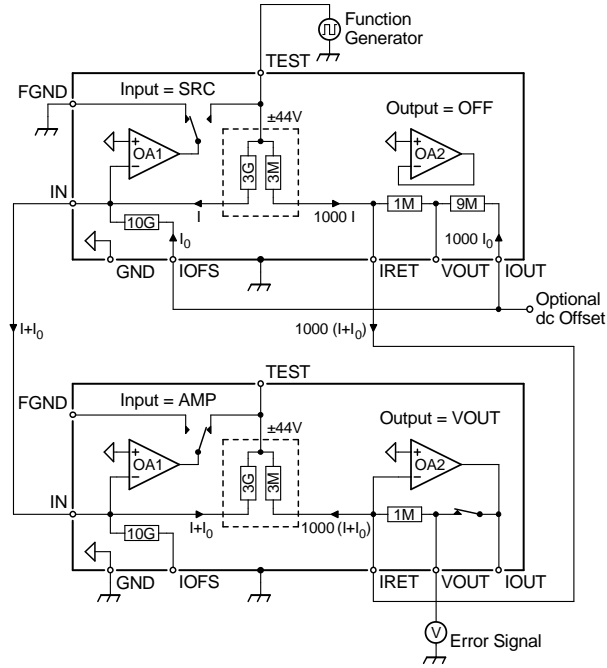

**Figure 4.** ULCA self-test configuration for comparing the resistor networks of two ULCA channels with each other. The current-generating ULCA is set into source mode with deactivated output stage (output = OFF). It provides the test currents for the active ULCA set into amplifier mode with voltage output. The error signal is measured at the VOUT connector. Alternatively, to maximize the error signal, current output with $R_{ext}$ = 100 MΩ can be used. An optional dc offset may be applied to investigate the linearity of the resistor networks.

Figure 5 shows a more detailed schematic of the ULCA with all features included. In addition to the optional input offset current (connector IOFS), an offset voltage may be applied via the VOFS connector. This way a bias voltage can be applied to the DUT, e.g., to monitor the current-voltage characteristic of an SET device during cool-down. To suppress high-frequency interference, an *R-C* circuit (0.8 Ω and 50 nF) is connected between internal ground and housing. A series connection of ten anti-parallel diodes limits the voltage between internal ground and housing to a safe level when the GND connector is left open. All input/output terminals are overvoltage-protected by appropriate series resistors. For the temperature monitoring, the internal voltage reference (AD780 from Analog Devices) is applied that has an analog temperature output proportional to absolute temperature (PTAT). It is buffered and filtered, and made accessible to the user at the TEMP connector. The typical transfer coefficient is 1.9 mV/K; the actual value is determined during the ULCA's calibration.

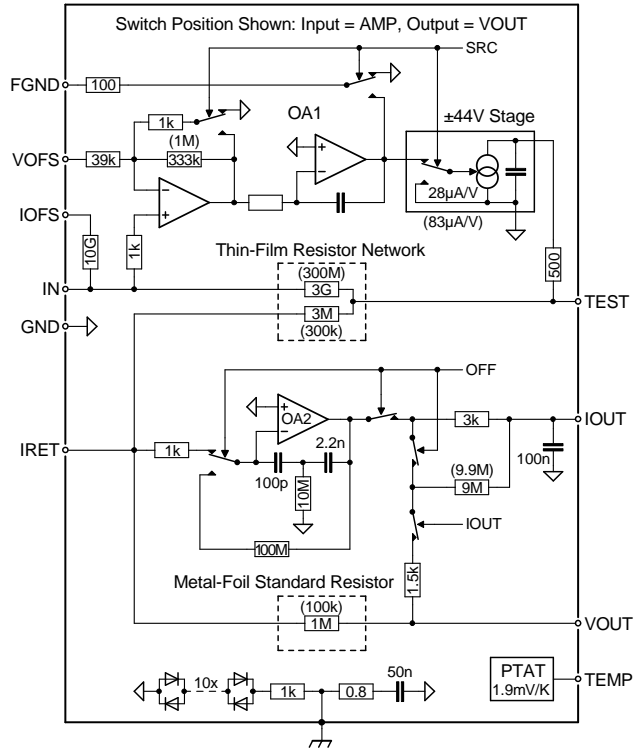

**Figure 5.** More detailed schematic of the ULCA. At dc the internal ground (open triangles) is isolated against the housing so that the reference potential can be defined via the GND input. Resistor values are shown for the low-noise ULCA variant (values in parentheses for the high-precision variant). To minimize interference, the circuit is battery powered.

*2.4. Traceable calibration*

Besides the very low current noise due to the high-ohmic input resistor network, ultra-high stability and traceable calibration of the transfer coefficient are key features of the ULCA. The traceable calibration is performed with a state-of-the art CCC resistance bridge developed in collaboration between PTB and the company Magnicon [9, 14]. The suitability of this special setup has been experimentally proven (see chapter 3), but other CCC resistance bridges may alternatively be used if the required turns ratios are available. The input and output stages of the ULCA are calibrated separately with different setups depicted in figure 6(*a*) and 6(*b*), respectively. The CCC resistance bridge basically consists of two highly-isolated current sources ($I_1$ and $I_2$) and a low-noise bridge-voltage detector (amplifier labeled nV). The currents $I_1$ and $I_2$ flow through the respective CCC windings $N_1$ and $N_2$ that are coupled to a SQUID. A so-called external feedback loop is realized that keeps the flux in the SQUID constant by controlling one of the current sources [9]. The constant dc flux offset is suppressed by applying current reversal. This way, the current ratio is made exactly equal to the inverse turns ratio, $I_1/I_2 = N_2/N_1$. To enable arbitrary current ratios and, consequently, arbitrary resistance ratios, a binary compensation network is added that drives an auxiliary winding (not shown in figure 6 for clarity) [15].

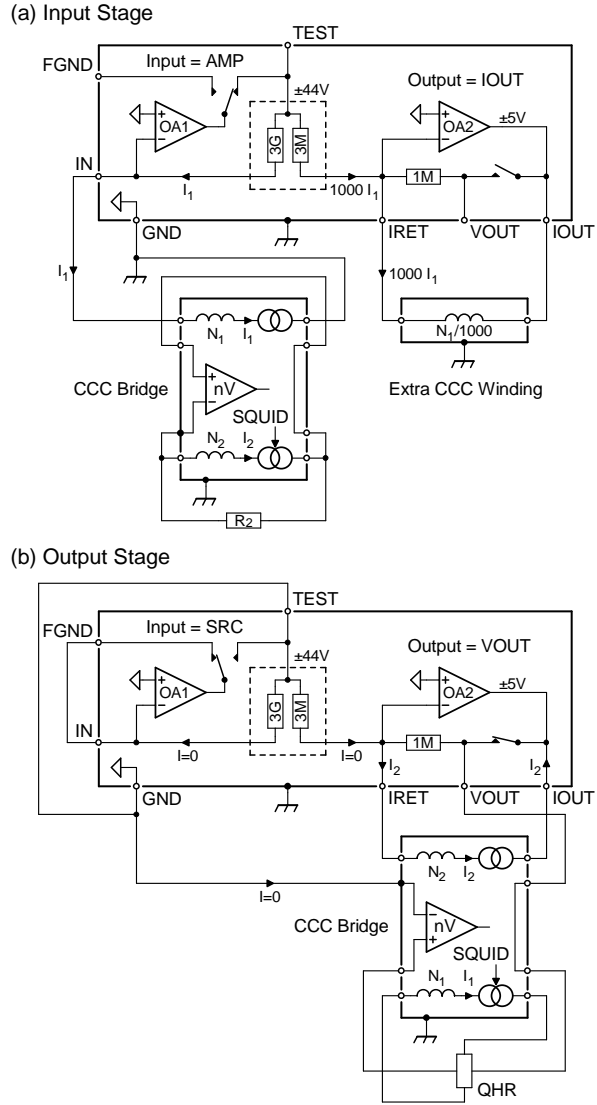

**Figure 6.** Calibration of the ULCA with a CCC. (*a*) $G_I$ of input stage, (*b*) $R_{IV}$ of output stage. The CCC resistance bridge involves two current sources $I_1$ and $I_2$. A vertical arrow indicates which current source is controlled by the external feedback loop of the SQUID. The negative input of the nanovoltmeter is connected to chassis (solid circle). In (*a*), an extra CCC winding with a matched number of turns is used to cancel the flux produced by the current $I_1$. The resistance bridge measures the gain error via the voltage drop across a calibrated resistor $R_2$. In (*b*) $R_{IV}$ is directly compared against the quantum Hall resistance or, optionally, a traceably calibrated standard resistor. The connection between FGND and IN provides feedback for OA1 ensuring adequate operating conditions.

The transfer coefficient of the input stage, the current gain $G_I = 1000$, is a dimensionless quantity. Therefore it is calibrated by using a pair of CCC windings with a 1000:1 turns ratio, i.e., $N_1 = 4000$ for the 12-bit CCC [9] and $N_1 = 16000$ for the 14-bit CCC [8]. The current $I_1$ is passed into GND flowing back through the ULCA input and the $N_1$ winding; the resulting 1000-fold current at the ULCA's output stage is fed through a winding $N_1/1000$ labeled in figure 6(*a*) as "extra CCC winding" because it is not required in case of normal resistance comparisons. The external feedback loop controls current source $I_2$. The feedback current is proportional to the deviation of the current gain $G_I$ from the nominal value of 1000. It flows through a standard resistor $R_2$, and the resulting voltage drop is measured with the bridge voltage detector. The demands on the accuracy of standard resistor $R_2$ are quite modest because the error signal is kept low (<1 ppm typically) by the binary compensation unit [15]. In case of a larger relative gain deviation, the turns ratio is adapted appropriately to relax the demands on the compensation unit.

The configuration for the calibration of the transfer coefficient of the output stage, $R_{IV} = 1$ MΩ, is depicted in figure 6(*b*). It resembles a conventional resistance comparison between a 1 MΩ standard resistor and the QHR, except that the 1 MΩ resistor is replaced by the ULCA output stage that appears as a virtual 1 MΩ resistor in a special four-terminal configuration. The current $I_2$ is passed into the ULCA output stage (IOUT, IRET) and the resulting voltage is tapped between the VOUT and GND connectors. The binary compensation network is adjusted such that the bridge signal is minimized. The residual bridge signal is measured with the bridge voltage detector and used to determine the resistance ratio. This way, $R_{IV}$ is traced back to the QHR. If a QHR is not available, a traceably calibrated standard resistor may be used instead.

*2.5.    Resistance calibration with ULCA*

The basic ULCA function is a 1000-fold amplification of a small electric current followed by a traceable conversion of the amplified current into a proportional voltage, either via the internal 1 MΩ resistor or by using an external standard resistor. Besides the traceable measurement or generation of sub-nA currents, this allows the calibration of high-ohmic resistors with ultimate performance. The circuit in figure 7(*a*) utilizes the ULCA's internal 1 MΩ resistor to compare the resistor under calibration (DUT) against the ULCA's transresistance $A_{TR}$. At the output, the difference between both appears as the error signal, i.e., the demands on the voltmeter accuracy are quite modest as the transresistance is chosen to match the resistance of the DUT. The standard 3 GΩ ULCA is well suited for direct calibration of 1 GΩ resistors. The 300 MΩ ULCA will be equipped with an internal 100 kΩ resistor instead of the standard 1 MΩ, which allows the calibration of 100 MΩ resistors at highest accuracy. For a sufficiently stable 100 MΩ standard resistor, the ULCA calibration can be verified by a direct calibration with 14-bit CCC against the QHR [8]. Such a consistency check might help to validate the CCC/ULCA setups or to find previously unknown effects contributing to the measurement uncertainty (e.g., higher leakage currents or stronger settling effects than expected).

For other values of the DUT, an external reference resistor can be used according to figure 7(*b*). Here, the DUT is compared against the external standard resistor in a setup similar to the self-test configuration presented in section 2.3. The current-generating ULCA in figure 4 is replaced by the combination of DUT plus reference resistor that should have a nominal 1000:1 resistance ratio. The output is proportional to the resistance deviation, which reduces the demands on the voltmeter accuracy substantially. For low-ohmic reference resistors, lead resistances have to be considered as discussed in section 2.1. Before mounting a resistor network into a ULCA, it can be tested with this setup which is convenient for components inspection. If the gain of the ULCA in figure 7(*b*) is sufficiently well known, the network under test can be calibrated this way. On the other hand, if a calibrated network (e.g., a more stable 300 MΩ network) is available for current generation, the ULCA's current gain can be calibrated without a CCC.

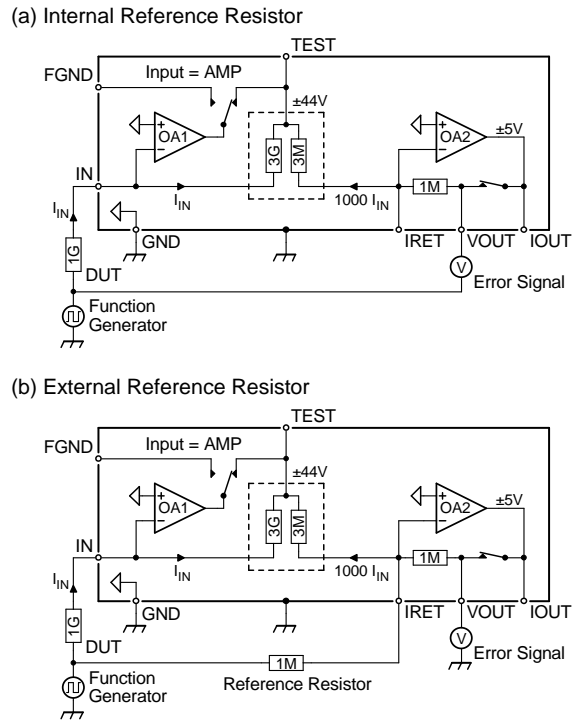

**Figure 7.** Calibration of high-ohmic resistors with the ULCA. (*a*) Internal reference resistor, (*b*) external reference resistor. In (*a*) the voltmeter measures the deviation between the resistance of the DUT and the transresistance $A_{TR}$ of the ULCA. The voltage across the DUT is limited by the ULCA output stage to ±5 V. In (*b*) the DUT is compared against the external reference resistor with a nominal 1000:1 resistance ratio. Here, the voltage across the DUT is limited by the ±44 V dynamic range of the ULCA input stage.

*2.6.    ULCA prototype development*

The initial ULCA concept [11] was developed at the beginning of 2012. By the end of 2012, a first two-channel prototype (ULCA-0) was realized. It is equipped with 1.4 GΩ SO-8 networks of the first design generation, yielding a noise level of 4 fA/√Hz in the frequency range 0.01-0.3 Hz. The ULCA-0 has a transresistance $A_{TR}$ = 50 MΩ that is calibrated in one step against the QHR or a traceably calibrated 12.9 kΩ standard resistor. A relatively high temperature coefficient of about +2 ppm/K was observed for both channels. The calibration was originally performed with PTB's 12-bit CCC [9]. After the 14-bit CCC [8] became available in September 2013, it was mostly used for the ULCA calibrations because of its lower current noise. During the first 1½ years of operation, the transresistance of the ULCA-0 remained within a span of about one part in $10^5$.

In 2013, the final ULCA concept with increased transresistance $A_{TR}$ = 1 GΩ and calibration in two steps was developed. Parallel to the improvement of the SO-8 resistor networks, the application of about 3000 identical 2 MΩ chip resistors for the 1000-fold current gain was introduced. For both variants it took several iterations to get the final circuit design.

In March 2014, a first single-channel ULCA-1 prototype became available This prototype is equipped with a homemade 3 GΩ 0805 network, i.e., the PCB was fabricated with an in-house circuit board plotter by milling and drilling, and the chip resistors were manually mounted and soldered. The instrument works well except that the input current gain $G_I$ shows an atypically high drift (-17 ppm/yr extrapolated from the first four months of operation). This is believed to be caused by leakage in the milled PCB and non-reproducible stresses in the chip resistors during hand-soldering. It was attempted to reduce the drift by annealing the network board for two hours at 153 °C. This treatment changed the relative gain deviation from –102 ppm to +13 ppm and the initial drift from about +0.09 ppm/d to -0.05 ppm/d, respectively. Note that this ULCA prototype was used for the study in [13] before the 3 GΩ network was annealed; the results in [12] were obtained with intermediate prototype variants that are not considered in this paper.

After performing the basic functional tests with the single-channel ULCA-1, a two-channel prototype was completed in June 2014 (see figure 2). This instrument is equipped with externally fabricated 3 GΩ resistor networks. In contrast to the homemade network, the copper on the PCBs is patterned by wet etch and the chip resistors are reflow soldered. As all resistors see the same temperature profile during soldering, a much better matching between the high-ohmic and low-ohmic sides and a correspondingly improved current gain stability are expected. Only one channel (CH B) is equipped with the current source option as it just came up during the final assembling of the two-channel unit.

## 3. Experimental performance

In this chapter, a comprehensive characterization of the above-mentioned single-channel ULCA-1 prototype is presented. A few initial measurement data from the two-channel ULCA-1 and a 300 MΩ test setup are also shown. If a result is obtained from an instrument other than the single-channel ULCA-1, it is explicitly mentioned; otherwise, the single-channel ULCA-1 is implied.

### 3.1. Input current noise

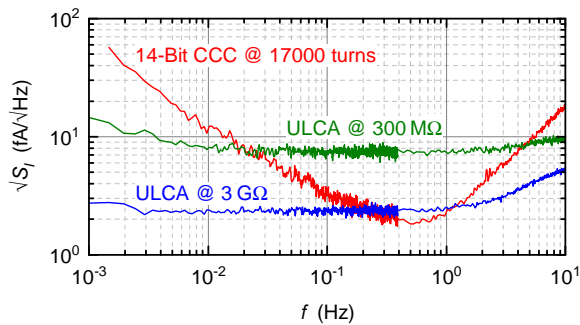

**Figure 8.** Input current noise spectra of two ULCA variants (3 GΩ and 300 MΩ) in comparison with the noise of the PTB 14-bit CCC [8]. The noise was measured with an HP35665A spectrum analyzer. For the ULCAs, current output with $R_{ext} = 100$ MΩ was used to maximize the output signal. The CCC noise was measured in a ratio-error test configuration. The current noise was calculated from the measured error voltage noise by using the known transfer coefficient of 0.65 nA/$\Phi_0$ for 17000 turns.

A low current noise is a prerequisite for acceptable averaging times at low input levels. The ULCA noise was investigated in the frequency range between 1 mHz and 10 Hz (figure 8). In addition to the 3 GΩ ULCA-1 prototype, a preliminary 300 MΩ setup was also tested. The amplifier input was open-circuited and covered by a connector cap to minimize power-line interference. The spectra were measured with a signal analyzer (35665A from Agilent) and the individual time traces were read out and stored for later analysis. Low white noise levels of 2.4 fA/√Hz and 7.5 fA/√Hz are achieved for the two variants, dominated by Nyquist noise in the resistor networks with a small contribution of <0.5 fA/√Hz from the input amplifier. Note that for floating current sources each terminal of the source may be connected to a separate input of a two-channel ULCA, thus measuring the current twice and increasing the signal-to-noise ratio by a factor of √2 [12]. Furthermore, correlation analysis between the two channels may provide extra information on noise sources in the setup.

At low frequencies, excess noise is commonly observed with a typical frequency dependence of the power spectral density $S \propto 1/f$. The $1/f$ corner is defined as the frequency at which the spectral densities of white noise and low-frequency excess noise are equal, corresponding to a factor of √2 increase in overall rms noise. The 3 GΩ ULCA shows an excellent low-frequency noise with a $1/f$ corner at about 1 mHz. This is substantially lower than for the 14-bit CCC wired as a current amplifier with 17000 input coil turns (cf. figure 8). The 300 MΩ variant has a higher $1/f$ noise with a corner frequency of about 2.5 mHz, still better than the CCC at frequencies below about 20 mHz. This is presumably caused by the input amplifier's voltage noise that appears as an extra noise current equal to the noise voltage divided by the network resistance. The low-frequency voltage noise could be suppressed by applying a chopper amplifier at the expense of additional switching noise (i.e., additional current noise) [16].

The Allan deviation is widely used for analyzing noise in low-level measurements. For white noise, it is equal to the standard deviation of the mean $\sigma_w$ [17]. It is related to the white noise spectral density $S_w$ by

$$\sigma_w = \sqrt{S_w / 2\tau} \quad , \tag{1}$$

where $\tau$ is the sampling time. Figure 9 shows Allan deviations calculated from the time traces that were used for the noise spectra in figure 8. The straight lines in figure 9 are calculated with (1) by using the white spectral densities determined from the spectra in figure 8. Excellent agreement between the Allan deviations and the straight lines is obtained at low sampling times $\tau$ where the noise is white. The 3 GΩ ULCA exhibits a minimum in the Allan deviation of about 0.1 fA at a sampling time near 1000 s, consistent with a $1/f$ corner frequency around 1 mHz. Note that the measurements were performed without temperature stabilization. The excellent low-frequency noise is made possible by the massive copper housing of the ULCA combined with a small and stable input bias current, that has a level of about 20 fA and a temperature coefficient of about 1 fA/K, respectively. This enables measurements with current reversal at very low repetition frequencies $f_R$ without impairing noise performance. For the 300 MΩ ULCA, the Allan deviation levels off at about 0.5 fA for sampling times above a few 100 s caused by $1/f$ amplifier noise [17].

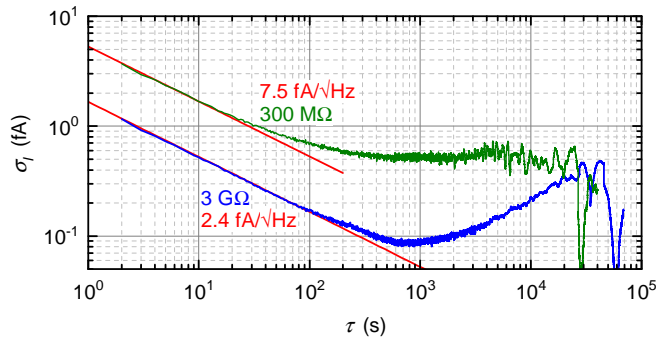

**Figure 9.** Allan deviation of the input current noise of the two ULCAs from figure 8. White noise is indicated by straight lines. The corresponding spectral densities are taken from the noise spectra in figure 8.

*3.2.     Settling after current reversal*

Generally, for precision dc measurements the signal to be measured is periodically reversed in order to suppress offset and drift effects. The amplitude of the resulting low-frequency square wave yields the desired dc level. In this paper, raw data analysis according to the method sketched in figure 7 of [15] is used to determine the result of dc current measurements with the ULCA. To suppress settling effects after current reversal, a certain number of data points is disregarded after each polarity change. However, beside the relatively uncritical exponential decays, slowly-decaying transients with $1/t$ dependence may occur that can falsify the result long after polarity reversal [14, 18]. Due to low-frequency excess noise it is not desirable to make the repetition frequency $f_R$ unnecessarily low. Therefore, an excellent settling after polarity reversal is crucial to obtain 0.1 ppm accuracy even with a relatively high repetition frequency $f_R$ = 0.05 Hz (typically used at PTB for CCC calibration).

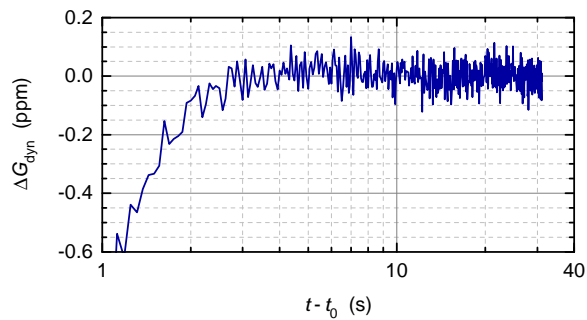

**Figure 10.** Settling behavior of CH B of the two-channel ULCA-1 depicted in figure 2. $\Delta G_{dyn}$ is the relative deviation of $G_I$ from its final value after decaying of transients, and $t-t_0$ is the time after current reversal. A total of 3260 responses were averaged to reduce noise. The currents into the IN and IRET inputs were generated by applying ±1.3 V to a 100 MΩ / 100 kΩ resistor pair similarly to the setup in figure 7(*b*). Transients in the current-generating resistor pair are negligible compared to the settling effects in the ULCA's 3 GΩ network.

The ULCA's resistor network was carefully optimized for minimum settling effects. By appropriate PCB layout, the secondary side of the network is used as a guard for the primary side. As much material as possible is milled away from the resistor board to minimize leakage and dielectric absorption effects. Careful mechanical design results in a still rigid construction. Guard rails on the edges of the resistor board eliminate leakage to the housing at the points of mechanical mounting. Special guard plates made from PCB are placed on top and bottom of the network board. The combination of all these measures yields a stable current gain and an excellent settling behavior. Figure 10 shows a settling to 0.1 ppm within about 2 s for a 3 GΩ resistor network. For comparison, the first 3 GΩ prototype showed pronounced distortions from dielectric absorption (tens of ppm after 2 s) and weak effects still visible minutes after current reversal. Note that the 300 MΩ variant should have a ten times lower dynamic deviation because of its ten times lower resistance as compared to the 3 GΩ network (assuming equal capacitance and dieelectric absorption for both types).

*3.3.     Calibration with CCC*

For the ULCA calibration, the currents through the CCC turns are periodically reversed, and the peak-peak amplitude of the resulting output signal is measured. Assuming 50% duty cycle, this means that the sampling time for each of the two current levels is half the total sampling time. In case of white noise this leads to a factor of √2 rise in standard deviation compared to (1). The standard deviation of the difference between the current levels increases by another factor of √2 because the variances of the two current levels are summed up. Therefore, the white noise uncertainty with polarity reversal $\sigma_{PR}$ is equal to $2\sigma_w$. Generalized one obtains

$$\sigma_{PR} = \sqrt{2S_e / \tau_e} \quad . \tag{2}$$

Here, $S_e$ is the effective white noise density with polarity reversal and $\tau_e$ is the effective sampling time without the intervals during which measurement data are disregarded to suppress settling effects after polarity reversal. $S_e$ and $S_w$ are equal for operation in the amplifier's white noise regime, i.e., if the repetition frequency $f_R$ is well above the $1/f$ corner of the amplifier noise. In general, $S_e \approx S(f_R)$ is found; an example is given in [8]. To obtain the relative uncertainty, $\sigma_{PR}$ has to be divided by the peak-peak value of the signal. For example, with $S_e = 2.4$ fA/√Hz and 20% data rejection, (2) yields $\sigma_{PR} = 20$ aA for a total measurement time of 10 h ($\tau_e = 8$ h), corresponding to a relative uncertainty of one part in $10^7$ for a current of ±100 pA (i.e., 200 pA peak-peak value). Note that the relative uncertainty is doubled if the current is switched on and off instead of being reversed because of the halved peak-peak value. For white amplifier noise and negligible settling effects (no disregarded data points), the same uncertainties are obtained in a given overall sampling time with and without polarity reversal.

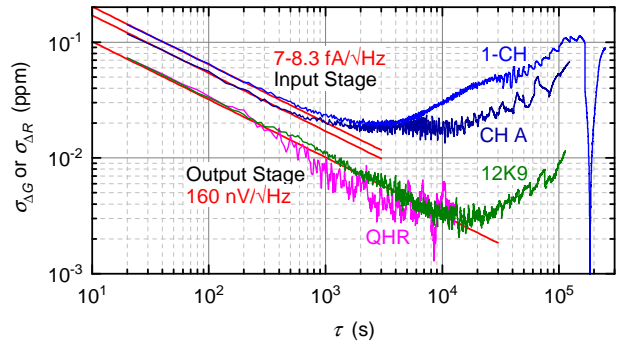

**Figure 11.** Allan deviations $\sigma_{\Delta G}$ (input stage) and $\sigma_{\Delta R}$ (output stage) of the relative deviations of the ULCA transfer coefficients measured with the 14-bit CCC. The output stage was calibrated either directly against the QHR or against a 12.9 kΩ wire-wound standard resistor (curves labeled QHR and 12K9). For the input stage, in addition to the single-channel ULCA-1, the result from CH A of the two-channel unit is also depicted (curves labeled 1-CH and CH A). The ULCA to be calibrated was placed in a temperature-stabilized air bath. The current was reversed every 10 s, and the first 5 s after each reversal were disregarded to reject transients. The 3 GΩ input stage was calibrated at ±13 nA, and the 1 MΩ output stage at ±0.5 V (limited by the allowed current through the QHR). For short measurement times $\tau$, white noise is observed (straight lines). The white noise levels are dominated by CCC current noise (input stage) or Nyquist noise in the 1 MΩ resistor plus amplifier noise (output stage), respectively.

Equation (2) is routinely used to check measurements with polarity reversal for excess noise. Figure 11 depicts Allan deviations obtained from the calibration of ULCA input and output stages. The corresponding white noise levels are

<8.3 fA/√Hz and 160 nV/√Hz, respectively, in good agreement with the values expected from CCC current noise, Nyquist noise in the 1 MΩ feedback resistor plus voltage and current noise from the ULCA's operational amplifiers. The calibration was performed with the PTB 14-bit CCC; alternatively, a 12-bit CCC may be used at the expense of a 4-fold rms current noise and a 4-fold uncertainty contribution from down-mixing of rf interference (SQUID nonlinearity) [8].

To illustrate the stability of the transfer coefficient over several days, the raw data of the longest input stage calibration (curve labeled 1-CH in figure 11) is depicted in figure 12 together with a smoothed curve. The peak-peak noise of the raw data increases after about 80 hours. The excess noise may be caused by external interference or random movement of trapped magnetic flux in the CCC or SQUID. It leads to the difference in the noise levels of the curves labeled 1-CH and CH A in figure 11. Note that a noise level below 10 fA/√Hz is more than adequate for the calibration of ULCA input stages. The smoothed curve shows peak-peak fluctuations of about 0.5 ppm. There is no drift visible in the displayed period of 139 h. Longer measurements were not possible due to the high degree of utilization of the CCC resistance bridge. Long-term investigations of the current gain stability over 6-8 weeks were done in a self-test configuration (see section 3.7).

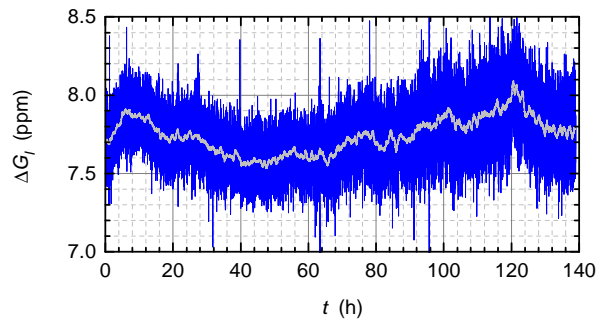

**Figure 12.** Stability of the relative current gain deviation $\Delta G_I$ of the input stage during a long CCC calibration. The raw data used for the curve labeled 1-CH in figure 11 is shown together with a moving average over 101 data points ($\tau = 2020$ s).

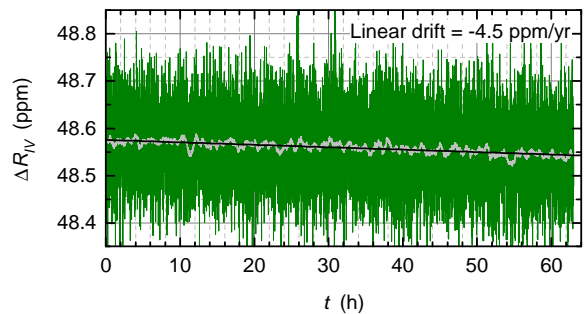

**Figure 13.** Stability of the relative transresistance deviation $\Delta R_{IV}$ of the output stage during a long CCC calibration. The raw data used for the curve labeled 12K9 in figure 11 is shown together with a moving average over 101 data points ($\tau = 2020$ s) and a straight line fit illustrating the drift.

The raw data of the longest output stage calibration (curve labeled 12K9 in figure 11) is displayed in figure 13. The peak-peak fluctuations are considerably smaller than for the input stage. Furthermore, a clear drift is observable even within the depicted period of 63 h. Surprisingly, the extrapolated drift of -4.5 ppm/yr for this short period is in fair agreement with the average drift of -3.4 ppm/yr found within the first four months of operation. Note that the 12.9 kΩ wire-wound standard resistor used for this calibration exhibits a negligibly small drift of about +0.015 ppm/yr (measured over a period of 20 years).

*3.4.    Uncertainty analysis*

In this section, the uncertainty of the calibration with CCC is discussed. Table 1 summarizes the uncertainty contributions of the input stage calibration. The noise component includes noise from the bridge voltage detector and short-term fluctuations of the current gain that cause the minimum observed in the Allan deviation. For the settling component, slowly decaying transients with a time dependence near $1/t$ are relevant only [14, 18]; exponential terms are negligible because the respective time constants are low and the first 5 s after each current reversal are disregarded (time counted relative to the start point of the 0.2 s current ramp). As a limit, a $1/t$ dependence with a maximum deviation of 0.1 ppm after 2 s is assumed (cf. figure 10) resulting in an uncertainty contribution of 0.016 ppm. The effect of SQUID nonlinearity is determined for a zero-flux deviation $\Phi_{NL}$ equally distributed in the range between $\pm 10^{-6}\,\Phi_0$. The CCC bridge contribution summarizes all systematic uncertainties related to the experimental setup, e.g., the gain error of the bridge voltage detector and the ratio errors of the CCC windings and the binary compensation unit [15].

**Table 1.** Uncertainty budget for the calibration of the 3 GΩ input stage. Unless otherwise noted, the quoted uncertainties are standard uncertainties (coverage factor $k = 1$) obtained with PTB's 14-bit CCC [8]; values for the 12-bit CCC [9] are given in parentheses if they differ from those for the 14-bit CCC.

| Component | Distribution | Comment | Uncertainty (ppm) |
|---|---|---|---|
| Noise | Normal | $\tau \approx 3600$ s | 0.03 (0.045) |
| Settling | Rectangular | $1/t$ dependence with <0.1 ppm after 2 s | 0.016 |
| SQUID nonlinearity | Rectangular | $|\Phi_{NL,max}| = 10^{-6}\,\Phi_0$ | 0.015 (0.06) |
| Resistor nonlinearity | Rectangular | <0.1 ppm/V | 0.047 |
| CCC bridge | Rectangular | Conservative limit | 0.003 |
| Total | | 95% confidence, $k = 2$ | 0.12 (0.18) |

Nonlinearity of the NiCr thin-film resistors has also to be considered (amplifier effects can be neglected because of the high open-loop gain $>10^9$). Typically, an upper limit for the voltage coefficient of 0.1 ppm/V is specified (probably the measurement limit of the resistor manufacturers). In table 1 this value is conservatively adopted, yielding an uncertainty contribution of 0.047 ppm for the voltage drop of 0.81 V across the 2 MΩ resistors of the low-ohmic side during calibration. This uncertainty contribution can presumably be considerably lowered after investigating the resistor nonlinearity at a level of a few parts in $10^8$ in a self-test setup, where the experiment can run over months to achieve a sufficiently low random uncertainty at 100 pA level. Corresponding measurements are under preparation.

For the ULCA calibration with 14-bit CCC, a total expanded uncertainty ($k$ = 2) of 12 parts in $10^8$ is achieved. With the 12-bit CCC, the uncertainty contributions from noise and SQUID nonlinearity rise (values in parentheses in table 1) because of the lower number of CCC turns (4000:4 instead of 16000:16). This increases the total expanded uncertainty to 18 parts in $10^8$. Note that the calibration uncertainty of the 300 MΩ input stage is expected to be substantially lower due to the improved resistor performance; appropriate calibration measurements are planned for the near future. The calibration uncertainty of the output stage is similar to that of a 1 MΩ standard resistor, yielding a total expanded uncertainty of about 2 parts in $10^8$ for the output stage both with the 12-bit and 14-bit CCC, respectively.

*3.5.    Temperature dependence*

The temperature dependence of the transfer coefficients of input and output stage was determined during initial calibration (figure 14). The ULCA was placed in a stabilized air bath and the temperature was varied between 22 °C and 25 °C at intervals of about 1 °C. A minimum waiting time of about 1½ hours was inserted after each change of the air bath's set point. Linear fits to the four data points (straight lines in figure 14) yield the respective temperature coefficients. For the temperature coefficient of the input stage, standard uncertainties of about 0.02 ppm/K (14-bit CCC) or 0.03 ppm/K (12-bit CCC), respectively, were estimated considering noise and short-term fluctuations of the resistance ratio. For the output stage, a conservative standard uncertainty of 0.01 ppm/K can be assumed independent of the CCC used for calibration.

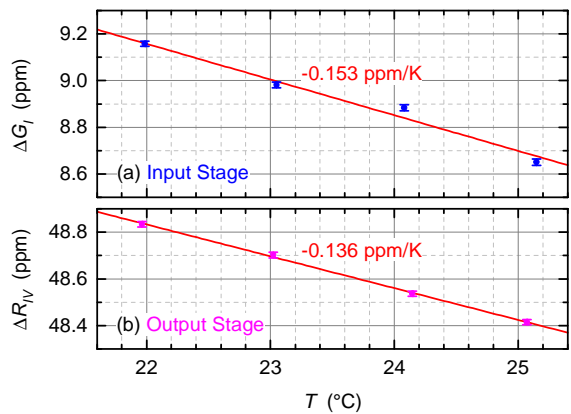

**Figure 14.** Temperature dependence of the ULCA transfer coefficients measured with the 14-bit CCC. (*a*) Relative current gain deviation $\Delta G_I$ of input stage, (*b*) relative transresistance deviation $\Delta R_{IV}$ of output stage directly compared against the QHR. The ULCA was placed in a temperature-stabilized air bath, and the TEMP output was used to determine the actual circuit temperature. About 1½ hours minimum waiting time were inserted after each change of the air bath's set temperature. Error bars indicate type A standard uncertainties ($k$ = 1).

Very low temperature coefficients of (–0.153±0.02) ppm/K and (–0.136±0.01) ppm/K are found for input and output stage of the single-channel ULCA-1 prototype in figure 14, respectively. Note that the temperature coefficients of the output stages of the two ULCA-1 prototypes agree within the standard uncertainty. These instruments are equipped with bulk metal-foil resistors from the same purchase order. The temperature coefficients of the input stages show a larger spread, (-0.146±0.02) ppm/K and (+0.01±0.02) ppm/K for CH A and CH B of the two-channel ULCA-1, respectively. The low temperature dependence of the transfer coefficients allows precise measurements even without extra temperature stabilization; the temperature stability of the laboratory air conditioner might be sufficient in most cases. For highest demands, the ULCA's built-in temperature sensor can be used to correct for temperature effects.

*3.6. Linearity*

The achievable calibration uncertainty of the input gain is directly affected by nonlinear effects in the resistor network. Therefore, the 14-bit CCC was used to experimentally evaluate an upper limit for the resistor nonlinearity. Linearity measurements are sensitive to short-term gain fluctuations due to the long averaging times required for achieving <0.1 ppm accuracy at low input currents. To minimize these effects, a special measurement sequence was applied in figures 15 and 16. Instead of measuring the data points of a diagram successively each with a long averaging time, short sampling times were used and the whole procedure was repeated until sufficient averaging effect was obtained for each point. This way, very long averaging times are possible without suffering from gain fluctuations, e.g., weeks or even months when using the self-test configuration described in section 2.3. With the 14-bit CCC, the measurement had been limited to about one day, yielding information at 0.1 ppm level only.

The gain nonlinearity $\Delta G_{NL}$ is defined here as the relative difference between the measured current gain and the mean current gain for all data point of the respective diagram. Two representative cases were investigated. Figure 15 shows the gain nonlinearity $\Delta G_{NL}$ versus the peak amplitude $I_P$ with zero offset current $I_0$ (i.e., the current was switched between $\pm I_P$), whereas figure 16 depicts the dependence of $\Delta G_{NL}$ on the offset current $I_0$ for fixed amplitude $I_P = 3.2$ nA. Within the random uncertainty, no clear dependence is visible at 0.1 ppm level, consistent with the typical limit of 0.1 ppm/V for thin-film resistors. Gain nonlinearity at substantially lower levels may be investigated with the self-test configuration (see section 2.3). Corresponding investigations are planned.

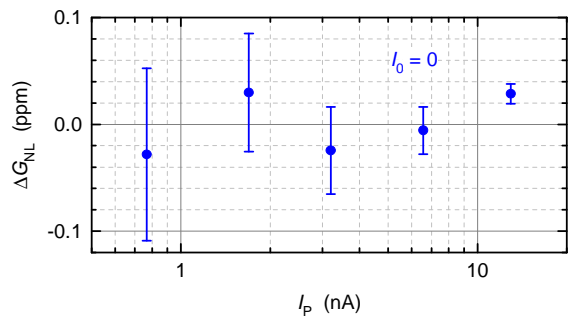

**Figure 15.** Nonlinearity $\Delta G_{NL}$ of the ULCA gain versus peak amplitude $I_P$ measured with the 14-bit CCC for zero offset $I_0 = 0$. The current was reversed every 10 s, and the first 2 s after each reversal were disregarded. Each data point was derived from five short measurements with 48 current reversals (96 reversals for $I_P = 1.7$ nA and 192 reversals for $I_P = 0.77$ nA). The total averaging time was 6 hours. Error bars indicate type A standard uncertainties ($k = 1$).

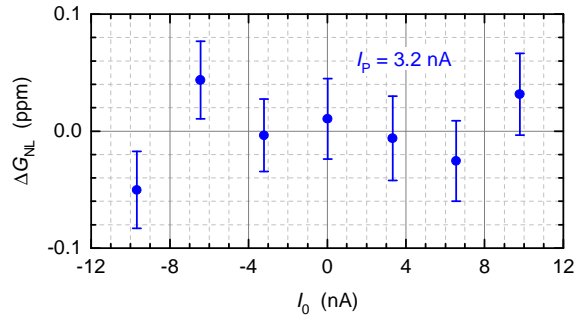

**Figure 16.** Nonlinearity $\Delta G_{NL}$ of the ULCA gain versus offset current $I_0$ measured with the 14-bit CCC for fixed peak amplitude $I_P$ = 3.2 nA. The current was reversed every 10 s, and the first 2 s after each reversal were disregarded. Each data point was derived from six short measurements with 48 current reversals. The total averaging time was about 5½ hours. Error bars indicate type A standard uncertainties ($k$ = 1).

*3.7. Gain stability*

To demonstrate the ULCA's performance in a low-level current measurement, a long-term self-test experiment according to figure 7(*b*) was carried out with the single-channel ULCA-1 at ±100 pA input current (figure 17). The measurement was started 12 days after annealing the hand-soldered 3 GΩ network. The input current was generated with a 4.9 GΩ resistor network mounted in a separate copper housing. This network comprises 384 SO-8 matched resistor pairs. Laser-trimming was not applied to the integrated resistor pairs in order to avoid potential detrimental effects on the matching performance. For the actual fabrication run, the mean resistance is 27% above the nominal value of 10 MΩ / 10 kΩ.

To compensate the systematic, slightly positive temperature coefficient of the 4.9 GΩ network, up to 16 PT100 thin-film sensors in case size 0805 can be added in series to the low-ohmic side of the network. The positions for these optional PT100 are equally distributed on the network board. If less than 16 PT100 are required, solder bumps are used as shorts at the remaining positions. At 23 °C, each PT100 introduces a nominal contribution of 109 Ω with a temperature coefficient of +0.39 Ω/K. As the temperature coefficient of the 4.9 GΩ network was not known, a total of 12 PT100 were added prior to the measurement according the expected intrinsic value of about +1 ppm/K. The remaining overall temperature coefficient of the setup was corrected for with the help of the ULCA's internal temperature sensor. By minimizing the temperature feedthrough in the experimental data, a net coefficient of –0.5 ppm/K was determined. Together with the known temperature coefficient of the single-channel ULCA-1 from CCC calibration (figure 14) this implies an intrinsic temperature coefficient of +1.3 ppm/K for the 4.9 GΩ network without PT100. Note that in a self-test configuration, the contribution of the current-generating networks enters into the calculation with inverted sign.

A peak-peak noise of about 10 ppm is observed in figure 17 for the selected $f_R$ = 4 mHz ($\tau$ = 250 s) due to the low input current of ±100 pA. The gray line shows the dependence smoothed by calculating the moving average over 101 data points (50 on either side of a respective value) corresponding to $\tau$ = 7 h. The smoothed curve fluctuates slowly around a linear drift of -19 ppm/yr with peak deviations of about ±0.8 ppm. The drift in figure 17 is close to the value of -17 ppm/yr determined for the single-channel ULCA-1 during the first four months of operation. This suggests that the long-term stability of 4.9 GΩ SO-8 network (used for current generation) is much better than that of the ULCA's built-in 3 GΩ network with hand-soldered chip resistors. For the network with integrated matched pairs, hand-soldering is expected to be acceptable because the low-ohmic and high-ohmic resistors of each pair are thermally very well coupled by the common substrate. Furthermore, thermal stress of the thin-film structure during soldering is lower with SO-8 package than for a 0805 chip resistor where the heat is directly transferred to the active resistor area.

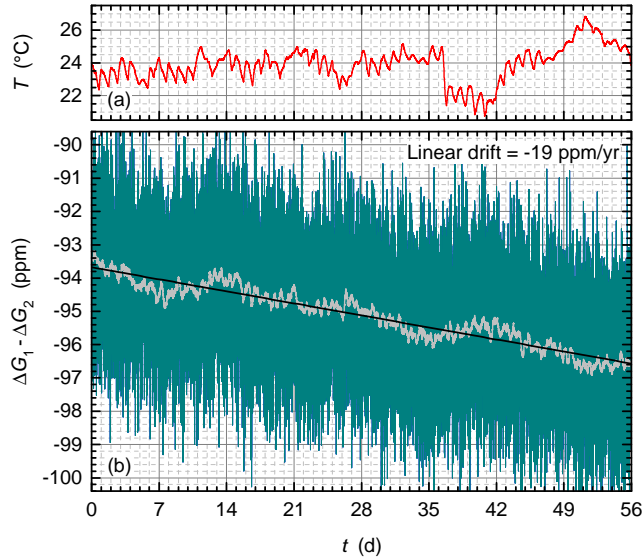

**Figure 17.** Stability of the ULCA gain measured in a self-test configuration. (*a*) Output of the internal temperature sensor, (*b*) difference between the relative current-gain deviation $\Delta G_1$ of the ULCA and that of a 4.9 GΩ SO-8 resistor network, $\Delta G_2$. The measurement was performed in an rf shielded room without temperature stabilization. The integrated temperature sensor was used to correct for the net temperature dependence of the resistor network combination (23 °C reference temperature). The current was reversed every 125 s, and the first 25 s after each reversal were disregarded to reject transients. Current output with $R_{ext}$ = 100 MΩ was used to maximize the ULCA's output voltage. A moving average over 101 data points ($\tau$ = 7 h) and a straight line fit illustrating the drift are depicted in (*b*).

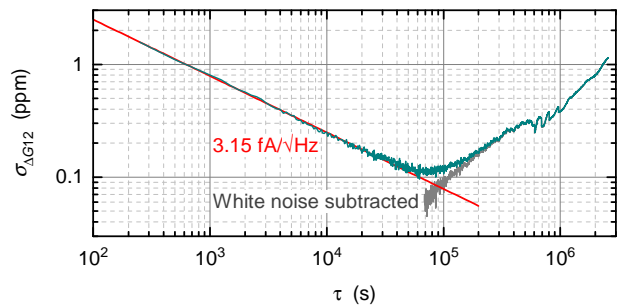

**Figure 18.** Allan deviation $\sigma_{\Delta G12}$ of the relative current-gain difference of the resistor network combination from figure 17. White noise is indicated by a straight line. The quoted noise level is consistent with Nyquist noise from the resistor networks. The gray curve shows an estimate for the gain fluctuations down to $\tau$ = 19 h obtained by subtracting the white noise contribution of the variance (straight line) from the total variance.

Figure 18 depicts the Allan deviation for the ±100 pA measurement from figure 17. A total white noise level of 3 fA/√Hz is expected at 23 °C due to Nyquist noise in the parallel connection of 3 GΩ and 4.9 GΩ resistors plus a small contribution from amplifier noise. The experimental white noise (straight line in figure 18) was determined by fitting (2) to the measured Allan deviation. It is only 5% above the theoretical white noise level, which is remarkable for the low repetition frequency $f_R$ = 4 mHz and shows the excellent low-frequency performance of the ULCA in a real low-level experiment. A minimum in the Allan deviation close to 0.1 ppm occurs at a sampling time $\tau$ of about one day. This demonstrates that the single-channel ULCA-1 is capable of performing measurements at a level of one part in $10^7$. The gray line in figure 18 illustrates the intrinsic gain fluctuations of the resistor network combination without the white noise contribution. It is consistent with the level observed for the ULCA's input stage calibration (curve labeled 1-CH in figure 11).

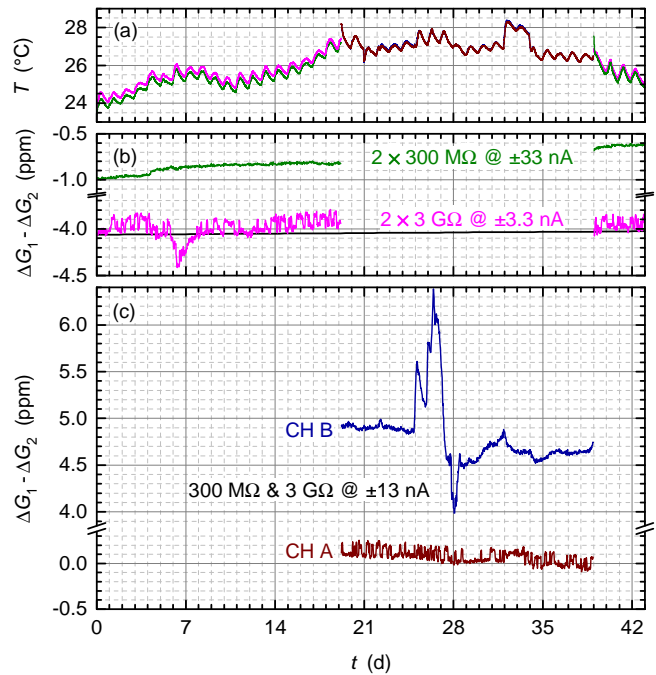

**Figure 19.** Stability of the ULCA gain measured in a self-test configuration. (*a*) Output of the internal temperature sensors, (*b*) and (*c*) difference between the relative current-gain deviations $\Delta G_1$ and $\Delta G_2$ of four resistor network combinations. During the first 19 and last 4 days (*b*), the two-channel 3 GΩ prototype depicted in figure 2 (CH A = AMP, CH B = SRC) and a single-channel 300 MΩ test variant with a 300 MΩ current-generating network were tested simultaneously. During the 20 days in between (*c*), the 300 MΩ networks were utilized to produce the test currents for the two channels of the 3 GΩ ULCA. The measurement was performed in an rf shielded room without temperature stabilization. The integrated temperature sensors were used to correct for the net temperature dependencies of the chosen resistor combinations (23 °C reference temperature). The current was reversed every 64 s, and the first 8 s after each reversal were disregarded to reject transients. In (*b*) and (*c*) each data point is the average of 16 subsequent current cycles ($\tau$ = 2048 s). The straight line indicates the expected value for the combination 2×3 GΩ obtained by linear interpolation between CCC calibrations before and after the experiment.

Figure 19 summarizes the initial gain stability characterization of the two-channel ULCA-1 and of the first 300 MΩ resistor networks. Two self-test setups were operated simultaneously in an environment without temperature stabilization. The temperature effect was considered by using the internal temperature sensors of the ULCAs and referencing the measured gain deviations to the nominal temperature of 23 °C. For the two-channel ULCA-1, the temperature coefficients from the CCC calibration were used (section 3.2). As the 300 MΩ networks had not yet been calibrated with a CCC, their temperature coefficients were estimated from the temperature-correlated fluctuations observed in the measured gain data during the experiment. The values used in figure 19 for temperature correction (+0.005 ppm/K and –0.025 ppm/K) confirm the superior quality of the individual 200 kΩ chip resistors as compared to the 2 MΩ parts of the 3 GΩ networks.

During the first period of 19 days, networks with equal resistance were compared with each other. The stability of the 3 GΩ networks was investigated by setting CH A of the two-channel ULCA-1 into AMP mode and CH B into SRC mode. For the 300 MΩ test setup, each resistor network was mounted into a separate copper housing. One network was cabled to an available ULCA prototype without resistor network housed in a separate copper housing, and the other was directly used for current generation. For the next 20 days, both ULCA-1 channels were set to the AMP mode, and the two 300 MΩ networks were cabled to generate the required test currents (the formerly current-generating for CH A, the other for CH B). During the last four days, the setup was switched back to the starting condition to check consistency and drift.

As expected, the 300 MΩ networks exhibit the highest stability. The 3 GΩ network of CH A shows a random two-level gain fluctuation that was also observed during the CCC calibration. Due to noise and short-term fluctuations, the separation between the levels cannot be determined precisely; a value of somewhat above 0.1 ppm can be estimated from figure 19. Nevertheless, over long periods excellent stability was observed. The 3 GΩ network of CH B generally exhibited good gain stability; however, during the second half of the fourth week large excursions of the output signal of CH B occurred over a few days. The gain returned to a value close to that before excursion and showed no more artifacts for the rest of the measurement. The reason for this temporary gain change is not yet clear. Figure 19 demonstrates that prior to demanding experiments in the sub-ppm level the resistor networks should be thoroughly checked, at least over a period of several weeks. The ULCA's built-in self-test option is a powerful method to do this with modest test equipment.

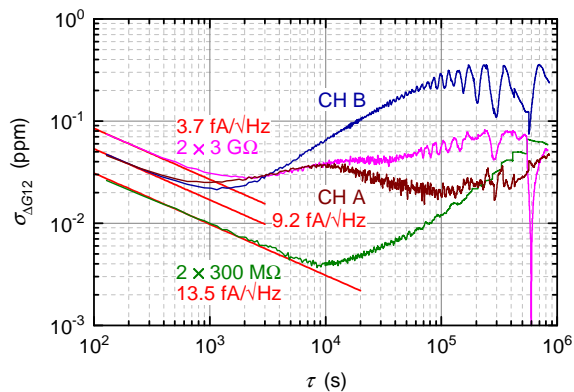

**Figure 20.** Allan deviation $\sigma_{\Delta G12}$ of the relative current-gain difference of the four resistor network combinations from figure 19 (the last four days of the measurement were not considered). White noise is indicated by straight lines. The corresponding levels result from Nyquist noise in the resistor networks and amplifier noise.

Figure 20 displays the Allan deviations for the measurement data from the four network combinations in figure 19. Again, the white noise levels deduced from the Allan deviations and those calculated from the circuit parameters agree well. For the 300 MΩ network combination, a very low minimum of 0.004 ppm at $\tau \approx 10^4$ s demonstrates excellent stability. For CH A of the 3 GΩ ULCA, a local *maximum* of about 0.04 ppm is observed at $\tau \approx 10^4$ s, caused by the two-level fluctuation. The Allan deviation from CH B is severely distorted by the large temporary gain fluctuation.

The two-channel ULCA-1 was calibrated with the 14-bit CCC within one week before and after the self-test experiment. The almost eight weeks interval in between allows an estimate of long-term drifts. The output stages of CH A and CH B showed nearly equal drift, -5.0 ppm/yr and -5.4 ppm/yr, respectively. These values are somewhat worse than the -3.4 ppm/yr observed for the single-channel ULCA-1 during the first four months of operation. For the input current gains of CH A and CH B, the change between the two calibrations was very small, only about -0.2 ppm. It was not sufficiently above the level of short-term gain fluctuations to provide a meaningful drift value. The long-term stability of the two externally fabricated 3 GΩ networks is substantially better than that of the homemade. The straight line in figure 19(*b*) shows the current gain calculated from the CCC calibrations for the 2×3 GΩ case under the assumption of linear drift between the two calibrations. Good agreement is found, in particular considering that the CCC calibration was done at the PTB site in Braunschweig and the self-test experiment in Berlin. Thus it appears that even the transfers between Berlin and Braunschweig did not noticeably affect the current gain. Furthermore, it demonstrates that the self-test setup yields results consistent with CCC calibration even though the self-test was performed in a laboratory with strong temperature fluctuations, whereas the calibration was done with the ULCA placed in a temperature-stabilized air bath.

Except for the severe instability of CH B during the second half of the fourth week in figure 19(*b*), the two-channel ULCA-1 features very high stability. There was only one artifact at such high level observed so far (i.e., during the initial two months of use after assembling). During the first 19 days in figure 19(*b*) (comparison of networks with equal resistance) the 3 GΩ network pair showed significantly higher gain fluctuations than the 300 MΩ pair. Therefore, we conclude that the problem was presumably caused by the 3 GΩ network of CH B. The stability of both ULCA-1 prototypes will be observed in future by regular calibrations. Experiments with the improved SO-8 resistor networks are also in preparation to evaluate which technology (individual chip resistors or integrated resistor pairs) exhibits the best long-term stability.

## 4. Conclusions and outlook

The ULCA is a powerful and flexible tool for the measurement or generation of small electric currents. It can be calibrated traceably to the quantum Hall effect and features an excellent stability of its transresistance $A_{TR}$. The ULCA provides a hitherto unrivalled precision and accuracy, even when compared to the best sub-nA current measurement method reported so far [3]. The precision of commercial instruments is outperformed by about two orders of magnitude [5]. Compared to current amplification with CCC, the ULCA offers better noise at low frequencies and avoids the systematic uncertainties that might appear at low currents due to mixing effects in the SQUID [8].

The ULCA current source may be an attractive alternative to the capacitor charging method [19] commonly used by national metrology institutes for calibrations in the sub-nA regime [5]. Here, the ULCA concept avoids the dominant uncertainty contribution from the frequency dependence of the capacitance [13] and allows the generation of arbitrary wave forms. Another potential ULCA application is the calibration of high-ohmic resistors. In this application, if the ULCA is combined with a conventional CCC bridge [9] for calibration, a performance is expected that is comparable to CCC bridges specially designed for high-ohmic resistors of up to 1 GΩ [20, 21].

The superior ULCA performance not only offers benefits for calibration applications, but also has implications for fundamental metrology research. For the development of SET devices generating relatively "high" currents of the order of 100 pA it becomes increasingly important to precisely measure their current-voltage characteristics in order to predict the achievable uncertainty on the basis of theoretical models (see for example [22]). The handy ULCA is well suited for such measurements, allowing ppm resolution at current levels of 100 pA with acceptable averaging times.

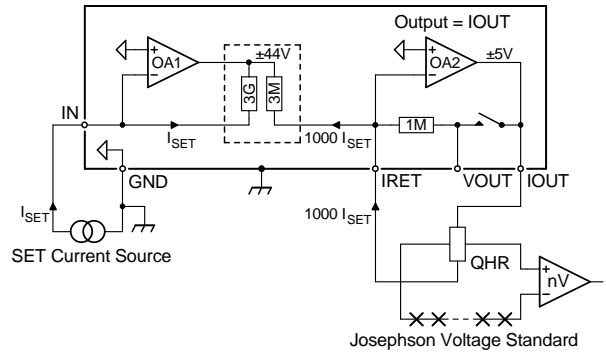

**Figure 21.** Direct quantum metrology triangle experiment with the ULCA amplifying the SET-generated current. The nominally 1000-fold current gain is calibrated with a CCC according to figure 6(*a*). The positive input of the nanovoltmeter is connected to the low-voltage side (near ground potential) of the QHR. For simplicity, the screens around SET current source, QHR, Josephson voltage standard and nanovoltmeter are omitted.

Finally, the ULCA is also a good candidate for a direct quantum metrology triangle experiment [23]. In this experiment, the current from an SET device would be measured with a ULCA in the current output mode, i.e, with the QHR directly used for $R_{ext}$ (see figure 22). The voltage across the QHR, 1.29 mV for 100 pA from the SET device, is compared with a matched voltage from a Josephson voltage standard using a nanovoltmeter as a null detector (for example [24]). In this configuration, the ULCA basically represents a 1000:1 CCC with superior low-frequency noise performance. Short-term drifts in the current gain $G_I$ can be suppressed by using two separate ULCAs and alternately measuring the SET current with one ULCA while the other is calibrated with a CCC. This way, the total measurement time can be extended considerably without increasing the uncertainty contribution due to the limited stability of the ULCA's current gain. Within a day, an uncertainty of about one part in $10^7$ realistically seems achievable at 100 pA direct current from the SET device.

**Acknowledgements**

The authors are indebted to Michael Piepenhagen for PCB fabrication and Monique Klemm for careful soldering of the 3 GΩ prototype network. Thanks also go to Martin Götz and Eckart Pesel for continuos help and support with the CCC measurements. Fruitful discussion with Bernhard Smandek on the ULCA's current source option is appreciated. This work was partially done within Joint Research Project "Qu-Ampere" (SIB07) supported by the European Metrology

Research Programme (EMRP). The EMRP is jointly funded by the EMRP participating countries within EURAMET and the European Union.